\documentclass[a4paper,11pt]{article}
\pdfoutput=1
\usepackage{pos}
\usepackage{tikz}    
\usetikzlibrary{graphs,graphs.standard,calc,automata}
\usepackage{tikz-feynman}
\tikzfeynmanset{compat=1.1.0}

\usepackage{caption}
\usepackage{subcaption}
\usepackage{graphicx}
\newtheorem{theorem}{Theorem}
\numberwithin{theorem}{section}

\newtheorem*{theorem*}{Theorem}

\newtheorem{corollary}[theorem]{Corollary}
\theoremstyle{definition}

\newtheorem{definition}[theorem]{Definition}

\theoremstyle{remark}

\newtheorem{example}[theorem]{Example}

\newcommand{\Gr}{\mathbf{Gr}}

\newcommand{\RP}{\mathbb{RP}}
\newcommand{\RR}{\mathbb{R}}

\newcommand{\QQ}{\mathbb{Q}}
\newcommand{\PP}{\mathbb{P}}

\newcommand{\CC}{\mathbb{C}}
\newcommand{\ZZ}{\mathbb{Z}}
\newcommand{\NN}{\mathbb{N}}

\newcommand{\U}{\mathcal{U}}
\newcommand{\F}{\mathcal{F}}
\newcommand{\G}{\mathcal{G}}
\newcommand{\cU}{\mathcal{U}}
\newcommand{\cF}{\mathcal{F}}
\newcommand{\cG}{\mathcal{G}}
\newcommand{\cI}{\mathcal{I}}
\newcommand{\cV}{\mathcal{V}}

\newcommand{\newt}{\mathbf{N}}
\newcommand{\conv}{\mathrm{conv}}

\newcommand{\xx}{\boldsymbol{x}}
\newcommand{\XX}{\boldsymbol{X}}
\newcommand{\soft}[1]{\texttt{#1}}
\newcommand{\ft}{\soft{feyntrop} }

\title{Tropical Feynman integration in the physical region}

\author[a]{Michael Borinsky}
\author[b]{Henrik J. Munch}
\author*[c]{Felix Tellander}

\affiliation[a]{Institute for Theoretical Studies,
ETH Zürich,
8092 Zürich, Switzerland}
\affiliation[b]{Dipartimento di Fisica e Astronomia,
Università degli Studi di Padova,
35131 Padova, Italy}

\affiliation[c]{Deutsches Elektronen-Synchrotron DESY, Notkestr.~85, 22607 Hamburg, Germany}

\emailAdd{michael.borinsky@eth-its.ethz.ch}
\emailAdd{henrikjessen.munch@studenti.unipd.it}
\emailAdd{felix@tellander.se}

\abstract{The software \ft for direct numerical evaluation of Feynman integrals is presented. We focus on the underlying combinatorics and polytopal geometries facilitating these methods. Especially matroids, generalized permutohedra and normality are discussed in detail.}

\FullConference{%
  EPS-HEP 2023\\
  21-25 Aug. 2023\\
  Hamburg, Germany
}

\begin{document}
\vspace*{-2\baselineskip}%
\hspace*{\fill} \mbox{\footnotesize{DESY-23-168}}
\maketitle

\section{Introduction}
The evaluation and understanding of Feynman integrals play an important role in many areas of modern physics, for example in particle accelerator phenomenology \cite{Heinrich:2020ybq}, gravitational wave physics \cite{Bern:2019nnu, Dlapa:2020cwj}, calculations of the magnetic moment of the muon \cite{Karshenboim:2005iy} and critical exponents in statistical field theory \cite{zinn2021quantum}. Many modern numerical methods uses the canonical differential equation approach \cite{Henn:2013pwa}, see e.g. \soft{AMFlow} \cite{Liu:2022chg}, \soft{DiffExp} \cite{Hidding:2020ytt} and \soft{SeaSyde} \cite{Armadillo:2022ugh}. Deriving the canonical differential equation is a potential bottleneck in these calculations which can be sidestepped using direct integration. Analytic integration can be performed in \soft{HyperInt} \cite{Panzer:2014caa} while numerical Monte Carlo techniques are for example implemented in \texttt{pySecDec} \cite{Borowka:2017idc}. The program \ft \cite{Borinsky:2023jdv} is of the latter type and integrates dimensionally regulated quasi-finite integrals numerically using Monte Carlo methods. The relevant sectors of integration are not determined using sector decomposition but relies on special properties of the Newton polytope of the integrand. If these polytopes are \emph{generalized permutohedra}, the decomposition into relevant sectors is greatly simplified. Integrable singularities of the integrand are regulated with the \emph{tropical approximation} \cite{Panzer:2019yxl} which also makes the integrand bounded from both above and below. For details on the tropical approximation, see \cite{Borinsky:2023jdv,Borinsky:2020rqs}. 

The program \ft has recently been used in \cite{Henn:2023vbd} to numerically verify the canonical differential equation result for a four-point three-loop process with one massive leg. It has also been used in \cite{Balduf:2023ilc} to calculate Feynman integrals in $\phi^4$-theory to 13 loops and beyond. The tropical way of thinking also sheds light on infrared singularities \cite{Arkani-Hamed:2022cqe} and how to calculate entire amplitudes directly without using Feynman integrals at all \cite{Arkani-Hamed:2023lbd}.

\section{Feynman Integrals and Generalized Hypergeometry}
Feynman integrals have many different equivalent representations, each with its own advantages and disadvantages. Consider one-particle irreducible Feynman graphs $G:=(E,V)$ with the number of cycles (loops) given by $L=|E|-|V|+1$. The vertex set $V$ has the disjoint partition $V=V_{\mathrm{ext}}\bigsqcup V_{\mathrm{int}}$ where each $u\in V_{\mathrm{ext}}$ is assigned an external incoming momenta $p_u\in\RR^{1,D-1}$. Each edge $e\in E$ is assigned a non-negative mass $m_e$.

For the purpose of direct numerical evaluation in \ft the following projective representation is used:
\begin{equation}\label{eq:par}
    \cI=\Gamma(\omega)\int_{\PP_+^{E}} \phi \quad
\text{ with }\quad
\phi = 
\left(\prod_{e\in E}\frac{x_e^{\nu_e}}{\Gamma(\nu_e)}\right)
\frac{1}{\cU(\xx)^{D/2}}\left(\frac{1}{\cV(\xx)-i\varepsilon \,\sum_{e\in E} x_e}\right)^{\omega}
\Omega\,.
\end{equation}
Where the integration domain is over the \emph{projective simplex}
$\PP_+^{E} = \{ \xx = [x_1: \cdots: x_{|E|}] \in \RP^{E-1} : x_e > 0 \}$
with respect to its canonical Kronecker form
\begin{align}
    \Omega=\sum_{e=1}^{|E|}(-1)^{|E|-e}\frac{d x_1}{x_1}\wedge\cdots\wedge\widehat{\frac{d x_e}{x_e}}\wedge\cdots\wedge\frac{d x_{|E|}}{x_{|E|}} \, .
\end{align}
The \emph{superficial degree of divergence} of the graph $G$ are 
given by $\omega = \sum_{e \in E} \nu_e - D L/2$. We write $\cV(\xx) = \cF(\xx)/\cU(\xx)$ as a shorthand for the quotient of the 
two \emph{Symanzik polynomials} that is defined from the underlying graph $G$ by
\begin{align}
\label{eq:polyUF}
    \cU(\xx)&:=\sum_{T} \prod_{e\notin T} x_e \, ,&
    \cF(\xx)&:=\cF_0+\cF_m=-\sum_{F} p(F)^2 \prod_{e \notin F}x_e+\cU(\xx)\sum_{e\in E}m_e^2x_e \, ,
\end{align}
where we sum over all spanning trees $T$ and all spanning two-forests $F$ of $G$, and $p(F)^2$ is the squared momentum running between the two-forest components. By this definition, $\cU$ and $\cF$ are homogeneous of degree $L$, resp. $L+1$, and hence $\cV$ is a homogeneous rational function of degree~$1$.
\subsection{Contour deformation}\label{sec: contour deformation}
In order to define the integral on the correct analytic branch we need to implement Feynman's causal $i\varepsilon$ prescription. This is done using a finite contour deformation respecting the projective invariance \cite{Hannesdottir:2022bmo}.

The deformation is given by the embedding $\iota_\lambda:\PP_+^{E}\hookrightarrow\CC\PP^{|E|-1}$:
\begin{equation}
    \iota_\lambda : 
x_e \mapsto X_e:=
x_e \exp \left (-i \lambda\frac{\partial  \mathcal V}{\partial x_e}(\xx) \right).
\end{equation}
Since  the boundary is characterized by $x_e=0$, $\iota_\lambda$ does not change the boundary. Using Cauchy's theorem, the integral is independent of $\iota_\lambda$ as long as the deformation does not cross any poles of $\phi$. Set
\begin{equation}
    \mathcal I = \Gamma(\omega) \int_{\iota_\lambda \left( \PP^{E}_+ \right)} \phi = 
\Gamma(\omega) \int_{\PP^{E}_+} \iota_\lambda^* \phi
\end{equation}
where $\iota_\lambda^*\phi$ is the pull-back, it can be written with the Jacobian $\iota_\lambda^*\Omega=\det(\mathcal{J}_\lambda(\xx))\Omega$ where
\begin{equation}
    \mathcal J_\lambda(\xx)^{e,h}
=\delta_{e,h} - i \lambda x_e \frac{\partial^2 \mathcal V}{\partial x_e \partial x_h} (\xx) \text{ for all } e,h \in E.
\end{equation}
 The deformed Feynman integral can thus be written as
\begin{equation}
     \cI=\Gamma(\omega)\int_{\PP_+^{E}} \iota^*_\lambda \, \phi= 
\Gamma(\omega)\int_{\PP_+^{E}} 
\left(\prod_{e\in E}\frac{X_e^{\nu_e} 
 }{\Gamma(\nu_e)}\right)
\frac{
\det \mathcal J_\lambda(\xx) 
}{\cU\left(\XX\right)^{D/2} 
\cdot
 \cV\left(\XX\right)^{\omega}}
\,
\Omega
\end{equation}
where $\XX=\iota_\lambda(\xx)$.

\subsection{Generalized hypergeometry}
Another useful representation is due to Lee and Pomeransky \cite{Lee2013}:  
\begin{equation}
\label{eq:I-def}
   \cI=\frac{\Gamma(D/2)}{\Gamma(D/2-\omega)}\int_0^\infty\left(\prod_{e \in E}\frac{x_e^{\nu_e}dx_e}{x_e\Gamma(\nu_e)}\right)\frac{1}{\G^{D/2}}\qquad \mathrm{where}\qquad\mathcal{G}=\mathcal{U}+\mathcal{F}.
\end{equation}
In this form, it is a generalized hypergeometric integral (Mellin transform) \cite{delaCruz:2019skx,Klausen:2019hrg} of the type studied by Passare and collaborators \cite{Nilsson2013,Berkesch2014}. This means that it satisfies a generalized hypergeometric system of partial differential equations in the sense of Gel'fand, Graev, Kapranov and Zelevinski\u{\i} (GGKZ, commonly shortened to GKZ) \cite{Gelfand1986,Gelfand1987,Gelfand1989,Gelfand1990,Gelfand1993}. 

Using multi-index notation we may write the Lee-Pomeransky polynomial as $\G=\sum_{i=1}^rc_ix^{\alpha_i}$ with $c_i\neq 0$ and $\alpha_i\in \ZZ_{\ge 0}^{{|E|}}$ for all $i=1,\ldots,r$. We define the two matrices
\begin{align}
    A&:=\{1\}\times A_-=\begin{pmatrix}
        1&1&\cdots&1,\\
        \alpha_1&\alpha_2&\cdots&\alpha_r
    \end{pmatrix}\in\ZZ_{\ge 0}^{({|E|}+1)\times r}, \; {\rm and}\label{eq:AMatrix}\\
    \beta&:=\begin{pmatrix}
        -D/2, & -\nu_1 &, \ldots &, -\nu_{{|E|}}
    \end{pmatrix}^T\in\CC^{{|E|}+1},\label{eq:BetaVector}
\end{align}
from which we construct the GKZ hypergeometric system $H_A(\beta)$ as the sum of two ideals:
\begin{align}
    I_A&:=\left\langle\partial^u-\partial^v\ |\ u,v\in\ZZ_{\ge 0}^{r}\ \mathrm{s.t.}\ Au=Av\right\rangle, \;\;{\rm and}\label{eq:toricIdeal}\\
    Z_A(\beta)&:=\left\langle \Theta_i(c, \partial)\;|\; \Theta= A\cdot\begin{pmatrix}
        c_1\partial_1\\
        \vdots\\
        c_r\partial_r
    \end{pmatrix}-\beta\right\rangle.
\end{align}
The ideal $I_A$ is actually an ideal in the {\em commutative} polynomial ring $\QQ[\partial_1, \cdots,\partial_r]$, and as such has a finite generating set $I_A=\langle h_1, \dots h_\ell  \rangle$ with $h_i\in \QQ[\partial_1, \cdots,\partial_r]$. This ideal $I_A$ is a {\em toric ideal} and it gives the defining equations of the projective {\em toric variety} $$
X_A=\{ z\in \PP^{r-1} \; |\; h_1(z)=\cdots= h_\ell(z)=0 \}
$$
associated to the matrix $A$, see e.g.~\cite{eisenbud1996binomial}, \cite[II, Chapter~5]{gelfand2008discriminants}.

\section{Generalized Permutohedra}\label{sec: GP}

\begin{figure}
 \centering
    \begin{subfigure}[b]{0.2\textwidth}
        \centering
        \begin{tikzpicture}
            [vertex/.style={inner sep=1pt,circle,draw=green!25!black,fill=green!75!black,thick}]
            \node[vertex,label=\(1\)] at (0, 0)     {};
        \end{tikzpicture}
    \end{subfigure}
    \hspace{-1cm}
    \begin{subfigure}[b]{0.25\textwidth}
        \centering
        \begin{tikzpicture}
             [
	           scale=1,
	           back/.style={loosely dotted, thin},
	           edge/.style={color=blue!95!black, thick},
	           facet/.style={fill=blue!95!black,fill opacity=0.400000},
	           vertex/.style={inner sep=1pt,circle,draw=green!25!black,fill=green!75!black,thick}]
            
            \draw[edge] (0,1) -- (1,0);
            \node[vertex,label=\(12\)] at (0,1) {};
            \node[vertex,label=\(21\)] at (1,0) {};
        \end{tikzpicture}
    \end{subfigure}
    \hspace{-0.5cm}
    \begin{subfigure}[b]{0.25\textwidth}
        \centering
        \begin{tikzpicture}%
	       [x={(-0.707031cm, -0.408259cm)},
	       y={(0.707183cm, -0.408200cm)},
	       z={(0.000025cm, 0.816516cm)},
	       scale=0.9,
	       back/.style={loosely dotted, thin},
	       edge/.style={color=blue!95!black, thick},
	       facet/.style={fill=blue!95!black,fill opacity=0.400000},
	       vertex/.style={inner sep=1pt,circle,draw=green!25!black,fill=green!75!black,thick}]
%
%

\coordinate (3.00000, 2.00000, 1.00000) at (3.00000, 2.00000, 1.00000);
\coordinate (3.00000, 1.00000, 2.00000) at (3.00000, 1.00000, 2.00000);
\coordinate (2.00000, 3.00000, 1.00000) at (2.00000, 3.00000, 1.00000);
\coordinate (1.00000, 3.00000, 2.00000) at (1.00000, 3.00000, 2.00000);
\coordinate (1.00000, 2.00000, 3.00000) at (1.00000, 2.00000, 3.00000);
\coordinate (2.00000, 1.00000, 3.00000) at (2.00000, 1.00000, 3.00000);
\fill[facet] (2.00000, 1.00000, 3.00000) -- (3.00000, 1.00000, 2.00000) -- (3.00000, 2.00000, 1.00000) -- (2.00000, 3.00000, 1.00000) -- (1.00000, 3.00000, 2.00000) -- (1.00000, 2.00000, 3.00000) -- cycle {};
\draw[edge] (3.00000, 2.00000, 1.00000) -- (3.00000, 1.00000, 2.00000);
\draw[edge] (3.00000, 2.00000, 1.00000) -- (2.00000, 3.00000, 1.00000);
\draw[edge] (3.00000, 1.00000, 2.00000) -- (2.00000, 1.00000, 3.00000);
\draw[edge] (2.00000, 3.00000, 1.00000) -- (1.00000, 3.00000, 2.00000);
\draw[edge] (1.00000, 3.00000, 2.00000) -- (1.00000, 2.00000, 3.00000);
\draw[edge] (1.00000, 2.00000, 3.00000) -- (2.00000, 1.00000, 3.00000);
\node[vertex,label=180:\(231\)] at (3.00000, 2.00000, 1.00000)     {};
\node[vertex,label=0:\(213\)] at (3.00000, 1.00000, 2.00000)     {};
\node[vertex,,label=0:\(321\)] at (2.00000, 3.00000, 1.00000)     {};
\node[vertex,label=180:\(312\)] at (1.00000, 3.00000, 2.00000)     {};
\node[vertex,label=\(132\)] at (1.00000, 2.00000, 3.00000)     {};
\node[vertex,label=\(123\)] at (2.00000, 1.00000, 3.00000)     {};
\end{tikzpicture}
\end{subfigure}
\hspace{1cm}
\begin{subfigure}[b]{0.25\textwidth}
\begin{tikzpicture}%
	[x={(-0.960291cm, 0.142643cm)},
	y={(-0.278999cm, -0.491199cm)},
	z={(-0.000076cm, 0.859288cm)},
	scale=0.8,
	back/.style={loosely dotted, thin},
	edge/.style={color=blue!95!black, thick},
	facet/.style={fill=blue!95!black,fill opacity=0.400000},
	vertex/.style={inner sep=1pt,circle,draw=green!25!black,fill=green!75!black,thick}]
%
%

\coordinate (-2.00000, -1.00000, 0.00000) at (-2.00000, -1.00000, 0.00000);
\coordinate (-2.00000, 0.00000, -1.00000) at (-2.00000, 0.00000, -1.00000);
\coordinate (-2.00000, 0.00000, 1.00000) at (-2.00000, 0.00000, 1.00000);
\coordinate (-2.00000, 1.00000, 0.00000) at (-2.00000, 1.00000, 0.00000);
\coordinate (-1.00000, -2.00000, 0.00000) at (-1.00000, -2.00000, 0.00000);
\coordinate (-1.00000, 0.00000, -2.00000) at (-1.00000, 0.00000, -2.00000);
\coordinate (-1.00000, 0.00000, 2.00000) at (-1.00000, 0.00000, 2.00000);
\coordinate (-1.00000, 2.00000, 0.00000) at (-1.00000, 2.00000, 0.00000);
\coordinate (0.00000, -2.00000, -1.00000) at (0.00000, -2.00000, -1.00000);
\coordinate (0.00000, -2.00000, 1.00000) at (0.00000, -2.00000, 1.00000);
\coordinate (0.00000, -1.00000, -2.00000) at (0.00000, -1.00000, -2.00000);
\coordinate (0.00000, -1.00000, 2.00000) at (0.00000, -1.00000, 2.00000);
\coordinate (0.00000, 1.00000, -2.00000) at (0.00000, 1.00000, -2.00000);
\coordinate (0.00000, 1.00000, 2.00000) at (0.00000, 1.00000, 2.00000);
\coordinate (0.00000, 2.00000, -1.00000) at (0.00000, 2.00000, -1.00000);
\coordinate (0.00000, 2.00000, 1.00000) at (0.00000, 2.00000, 1.00000);
\coordinate (1.00000, -2.00000, 0.00000) at (1.00000, -2.00000, 0.00000);
\coordinate (1.00000, 0.00000, -2.00000) at (1.00000, 0.00000, -2.00000);
\coordinate (1.00000, 0.00000, 2.00000) at (1.00000, 0.00000, 2.00000);
\coordinate (1.00000, 2.00000, 0.00000) at (1.00000, 2.00000, 0.00000);
\coordinate (2.00000, -1.00000, 0.00000) at (2.00000, -1.00000, 0.00000);
\coordinate (2.00000, 0.00000, -1.00000) at (2.00000, 0.00000, -1.00000);
\coordinate (2.00000, 0.00000, 1.00000) at (2.00000, 0.00000, 1.00000);
\coordinate (2.00000, 1.00000, 0.00000) at (2.00000, 1.00000, 0.00000);
\draw[edge,back] (-2.00000, -1.00000, 0.00000) -- (-1.00000, -2.00000, 0.00000);
\draw[edge,back] (-1.00000, -2.00000, 0.00000) -- (0.00000, -2.00000, -1.00000);
\draw[edge,back] (-1.00000, -2.00000, 0.00000) -- (0.00000, -2.00000, 1.00000);
\draw[edge,back] (-1.00000, 0.00000, -2.00000) -- (0.00000, -1.00000, -2.00000);
\draw[edge,back] (0.00000, -2.00000, -1.00000) -- (0.00000, -1.00000, -2.00000);
\draw[edge,back] (0.00000, -2.00000, -1.00000) -- (1.00000, -2.00000, 0.00000);
\draw[edge,back] (0.00000, -2.00000, 1.00000) -- (0.00000, -1.00000, 2.00000);
\draw[edge,back] (0.00000, -2.00000, 1.00000) -- (1.00000, -2.00000, 0.00000);
\draw[edge,back] (0.00000, -1.00000, -2.00000) -- (1.00000, 0.00000, -2.00000);
\draw[edge,back] (1.00000, -2.00000, 0.00000) -- (2.00000, -1.00000, 0.00000);
\draw[edge,back] (2.00000, -1.00000, 0.00000) -- (2.00000, 0.00000, -1.00000);
\draw[edge,back] (2.00000, -1.00000, 0.00000) -- (2.00000, 0.00000, 1.00000);
\node[vertex] at (2.00000, -1.00000, 0.00000)     {};
\node[vertex] at (-1.00000, -2.00000, 0.00000)     {};
\node[vertex] at (0.00000, -2.00000, -1.00000)     {};
\node[vertex] at (0.00000, -1.00000, -2.00000)     {};
\node[vertex] at (0.00000, -2.00000, 1.00000)     {};
\node[vertex] at (1.00000, -2.00000, 0.00000)     {};
\fill[facet] (2.00000, 1.00000, 0.00000) -- (1.00000, 2.00000, 0.00000) -- (0.00000, 2.00000, 1.00000) -- (0.00000, 1.00000, 2.00000) -- (1.00000, 0.00000, 2.00000) -- (2.00000, 0.00000, 1.00000) -- cycle {};
\fill[facet] (2.00000, 1.00000, 0.00000) -- (1.00000, 2.00000, 0.00000) -- (0.00000, 2.00000, -1.00000) -- (0.00000, 1.00000, -2.00000) -- (1.00000, 0.00000, -2.00000) -- (2.00000, 0.00000, -1.00000) -- cycle {};
\fill[facet] (-2.00000, 1.00000, 0.00000) -- (-2.00000, 0.00000, -1.00000) -- (-2.00000, -1.00000, 0.00000) -- (-2.00000, 0.00000, 1.00000) -- cycle {};
\fill[facet] (1.00000, 0.00000, 2.00000) -- (0.00000, -1.00000, 2.00000) -- (-1.00000, 0.00000, 2.00000) -- (0.00000, 1.00000, 2.00000) -- cycle {};
\fill[facet] (0.00000, 2.00000, -1.00000) -- (-1.00000, 2.00000, 0.00000) -- (-2.00000, 1.00000, 0.00000) -- (-2.00000, 0.00000, -1.00000) -- (-1.00000, 0.00000, -2.00000) -- (0.00000, 1.00000, -2.00000) -- cycle {};
\fill[facet] (0.00000, 2.00000, 1.00000) -- (-1.00000, 2.00000, 0.00000) -- (-2.00000, 1.00000, 0.00000) -- (-2.00000, 0.00000, 1.00000) -- (-1.00000, 0.00000, 2.00000) -- (0.00000, 1.00000, 2.00000) -- cycle {};
\fill[facet] (1.00000, 2.00000, 0.00000) -- (0.00000, 2.00000, -1.00000) -- (-1.00000, 2.00000, 0.00000) -- (0.00000, 2.00000, 1.00000) -- cycle {};
\draw[edge] (-2.00000, -1.00000, 0.00000) -- (-2.00000, 0.00000, -1.00000);
\draw[edge] (-2.00000, -1.00000, 0.00000) -- (-2.00000, 0.00000, 1.00000);
\draw[edge] (-2.00000, 0.00000, -1.00000) -- (-2.00000, 1.00000, 0.00000);
\draw[edge] (-2.00000, 0.00000, -1.00000) -- (-1.00000, 0.00000, -2.00000);
\draw[edge] (-2.00000, 0.00000, 1.00000) -- (-2.00000, 1.00000, 0.00000);
\draw[edge] (-2.00000, 0.00000, 1.00000) -- (-1.00000, 0.00000, 2.00000);
\draw[edge] (-2.00000, 1.00000, 0.00000) -- (-1.00000, 2.00000, 0.00000);
\draw[edge] (-1.00000, 0.00000, -2.00000) -- (0.00000, 1.00000, -2.00000);
\draw[edge] (-1.00000, 0.00000, 2.00000) -- (0.00000, -1.00000, 2.00000);
\draw[edge] (-1.00000, 0.00000, 2.00000) -- (0.00000, 1.00000, 2.00000);
\draw[edge] (-1.00000, 2.00000, 0.00000) -- (0.00000, 2.00000, -1.00000);
\draw[edge] (-1.00000, 2.00000, 0.00000) -- (0.00000, 2.00000, 1.00000);
\draw[edge] (0.00000, -1.00000, 2.00000) -- (1.00000, 0.00000, 2.00000);
\draw[edge] (0.00000, 1.00000, -2.00000) -- (0.00000, 2.00000, -1.00000);
\draw[edge] (0.00000, 1.00000, -2.00000) -- (1.00000, 0.00000, -2.00000);
\draw[edge] (0.00000, 1.00000, 2.00000) -- (0.00000, 2.00000, 1.00000);
\draw[edge] (0.00000, 1.00000, 2.00000) -- (1.00000, 0.00000, 2.00000);
\draw[edge] (0.00000, 2.00000, -1.00000) -- (1.00000, 2.00000, 0.00000);
\draw[edge] (0.00000, 2.00000, 1.00000) -- (1.00000, 2.00000, 0.00000);
\draw[edge] (1.00000, 0.00000, -2.00000) -- (2.00000, 0.00000, -1.00000);
\draw[edge] (1.00000, 0.00000, 2.00000) -- (2.00000, 0.00000, 1.00000);
\draw[edge] (1.00000, 2.00000, 0.00000) -- (2.00000, 1.00000, 0.00000);
\draw[edge] (2.00000, 0.00000, -1.00000) -- (2.00000, 1.00000, 0.00000);
\draw[edge] (2.00000, 0.00000, 1.00000) -- (2.00000, 1.00000, 0.00000);
\node[vertex] at (-2.00000, -1.00000, 0.00000)     {};
\node[vertex] at (-2.00000, 0.00000, -1.00000)     {};
\node[vertex] at (-2.00000, 0.00000, 1.00000)     {};
\node[vertex] at (-2.00000, 1.00000, 0.00000)     {};
\node[vertex] at (-1.00000, 0.00000, -2.00000)     {};
\node[vertex] at (-1.00000, 0.00000, 2.00000)     {};
\node[vertex] at (-1.00000, 2.00000, 0.00000)     {};
\node[vertex] at (0.00000, -1.00000, 2.00000)     {};
\node[vertex] at (0.00000, 1.00000, -2.00000)     {};
\node[vertex] at (0.00000, 1.00000, 2.00000)     {};
\node[vertex] at (0.00000, 2.00000, -1.00000)     {};
\node[vertex] at (0.00000, 2.00000, 1.00000)     {};
\node[vertex] at (1.00000, 0.00000, -2.00000)     {};
\node[vertex] at (1.00000, 0.00000, 2.00000)     {};
\node[vertex] at (1.00000, 2.00000, 0.00000)     {};
\node[vertex] at (2.00000, 0.00000, -1.00000)     {};
\node[vertex] at (2.00000, 0.00000, 1.00000)     {};
\node[vertex] at (2.00000, 1.00000, 0.00000)     {};
\end{tikzpicture}
\end{subfigure}
\caption{The first four permutohedra where it is easily seen from the explicit vertex coordinates (permutations) that all edges are parallel to $\mathbf{e}_i-\mathbf{e}_j$ for $i\neq j$.}
\label{fig: permutohedron}
\end{figure}
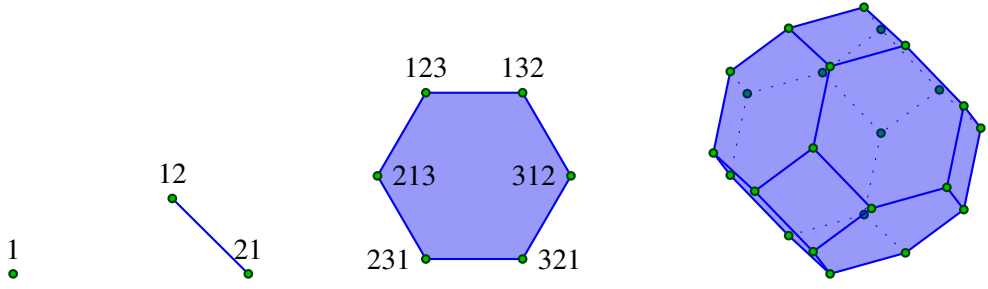

The classical \emph{perumtohedron} is a polytopal model of permutations, see Fig. \ref{fig: permutohedron}. For $n$ elements the permutohedron $P_n$ is the $(n-1)$-dimensional polytope in $\RR^n$ with vertices $(\sigma(1),\ldots,\sigma(n))$ where $\sigma$ runs over all permutations of $[n]:=\{1,2,\ldots,n\}$. Every point in $P_n$ satisfy $\sum_ix_i=n(n+1)/2$, meaning that $P_n$ lies in a hyperplane and hence $\mathrm{dim}(P_n)=n-1$. Note that every edge in $P_n$ is parallel to $\mathbf{e}_i-\mathbf{e}_j$ for some $i\neq j$ where $\mathbf{e}_i$ denote the standard basis of $\RR^n$.

    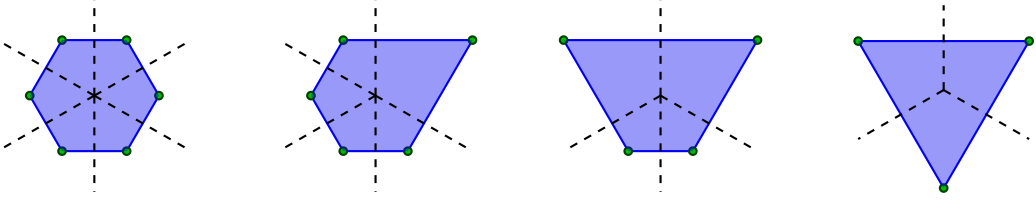
\begin{figure}
        \centering
        \begin{subfigure}{0.24\textwidth}
            \centering
            \begin{tikzpicture}
	           [scale=0.85,
	           normal/.style={dashed, thick},
	           edge/.style={color=blue!95!black, thick},
	           facet/.style={fill=blue!95!black,fill opacity=0.400000},
	           vertex/.style={inner sep=1pt,circle,draw=green!25!black,fill=green!75!black,thick}]
            \draw[edge](1,0)--(0.5,0.866)--(-0.5,0.866)--(-1,0)--(-0.5,-0.866)--(0.5,-0.866)--(1,0);
            \node[vertex] at (1,0) {};
            \node[vertex] at (0.5,0.866) {};
            \node[vertex] at (-0.5,0.866) {};
            \node[vertex] at (-1,0) {};
            \node[vertex] at (-0.5,-0.866) {};
            \node[vertex] at (0.5,-0.866) {};

            \fill[facet] (1,0)--(0.5,0.866)--(-0.5,0.866)--(-1,0)--(-0.5,-0.866)--(0.5,-0.866)--(1,0) {};

            \draw[normal] (0,0)--(0,1.5);
            \draw[normal] (0,0)--(0,-1.5);
            \draw[normal] (0,0)--(1.5,0.866);
            \draw[normal] (0,0)--(1.5,-0.866);
            \draw[normal] (0,0)--(-1.5,0.866);
            \draw[normal] (0,0)--(-1.5,-0.866);
            
            \end{tikzpicture}
        \end{subfigure}
        \begin{subfigure}{0.24\textwidth}
            \centering
            \begin{tikzpicture}
                [scale=0.85,
	           normal/.style={dashed, thick},
	           edge/.style={color=blue!95!black, thick},
	           facet/.style={fill=blue!95!black,fill opacity=0.400000},
	           vertex/.style={inner sep=1pt,circle,draw=green!25!black,fill=green!75!black,thick}]
            \draw[edge](1.5,0.866)--(-0.5,0.866)--(-1,0)--(-0.5,-0.866)--(0.5,-0.866)--(1.5,0.866);
            \node[vertex] at (1.5,0.866) {};
            \node[vertex] at (-0.5,0.866) {};
            \node[vertex] at (-1,0) {};
            \node[vertex] at (-0.5,-0.866) {};
            \node[vertex] at (0.5,-0.866) {};

            \fill[facet] (1.5,0.866)--(-0.5,0.866)--(-1,0)--(-0.5,-0.866)--(0.5,-0.866)--(1.5,0.866) {};

            \draw[normal] (0,0)--(0,1.5);
            \draw[normal] (0,0)--(0,-1.5);
            \draw[normal] (0,0)--(1.5,-0.866);
            \draw[normal] (0,0)--(-1.5,0.866);
            \draw[normal] (0,0)--(-1.5,-0.866);
            \end{tikzpicture}
        \end{subfigure}
        \begin{subfigure}{0.24\textwidth}
            \centering
            \begin{tikzpicture}
                [scale=0.85,
	           normal/.style={dashed, thick},
	           edge/.style={color=blue!95!black, thick},
	           facet/.style={fill=blue!95!black,fill opacity=0.400000},
	           vertex/.style={inner sep=1pt,circle,draw=green!25!black,fill=green!75!black,thick}]
            \draw[edge](1.5,0.866)--(-1.5,0.866)--(-0.5,-0.866)--(0.5,-0.866)--(1.5,0.866);
            \node[vertex] at (1.5,0.866) {};
            \node[vertex] at (-1.5,0.866) {};
            \node[vertex] at (-0.5,-0.866) {};
            \node[vertex] at (0.5,-0.866) {};

            \fill[facet] (1.5,0.866)--(-1.5,0.866)--(-0.5,-0.866)--(0.5,-0.866)--(1.5,0.866) {};

            \draw[normal] (0,0)--(0,1.5);
            \draw[normal] (0,0)--(0,-1.5);
            \draw[normal] (0,0)--(1.5,-0.866);
            \draw[normal] (0,0)--(-1.5,-0.866);
            \end{tikzpicture}
        \end{subfigure}
        \begin{subfigure}{0.24\textwidth}
            \centering
            \begin{tikzpicture}
                 [scale=0.75,
	           normal/.style={dashed, thick},
	           edge/.style={color=blue!95!black, thick},
	           facet/.style={fill=blue!95!black,fill opacity=0.400000},
	           vertex/.style={inner sep=1pt,circle,draw=green!25!black,fill=green!75!black,thick}]
            \draw[edge](1.5,0.866)--(-1.5,0.866)--(0,-2*0.866)--(1.5,0.866);
            \node[vertex] at (1.5,0.866) {};
            \node[vertex] at (-1.5,0.866) {};
            \node[vertex] at (0,-2*0.866) {};

            \fill[facet] (1.5,0.866)--(-1.5,0.866)--(0,-2*0.866)--(1.5,0.866) {};

            \draw[normal] (0,0)--(0,1.5);
            \draw[normal] (0,0)--(1.5,-0.866);
            \draw[normal] (0,0)--(-1.5,-0.866);
            \end{tikzpicture}
        \end{subfigure}
        \caption{A permutohedron and the three deformations of it into generalized permutohedra, keeping the edge directions parallel to $\mathbf{e}_i-\mathbf{e}_j$ for $i\neq j$.}
        \label{fig:GP}
    \end{figure}

In the theory of GKZ, the permutohedron appears as a \emph{secondary polytope} $\Sigma(A)$, where $A$ denotes the vertices of the triangular prism $Q=\Delta^1\times\Delta^{n-1}$. The vertices (or equivalently, the top-dimensional normal cones) of $\Sigma(A)$ correspond to regular triangularizations of $Q$ and also to regions of convergence of series solutions to the generalized hypergeometric system defined by $A$ \cite{gelfand2008discriminants}. 

As remarked above, the permutohedron has the property that all edges are parallel to $\mathbf{e}_i-\mathbf{e}_j$, this is the defining property of a \emph{generalized permutohedra} (GP) \cite{Postnikov2005} (cf. Fig. \ref{fig:GP}) but can also be strengthened to define the \emph{matroid polytope} \cite{GelfandSerganova,GGMS}.
\begin{definition}[Generalized permutohedron]
    A polytope $P$ is said to be a \emph{generalized permutohedron} if all its edges are parallel to $\mathbf{e}_i-\mathbf{e}_j$ for some $i\neq j$.
\end{definition}
\begin{definition}[Matroid polytope]
    A polytope $P$ is said to be a \emph{matroid polytope} if all its vertices lie in a hypersimplex and all edges are equal to $\mathbf{e}_i-\mathbf{e}_j$ for some $i\neq j$.
\end{definition}
We remark that a matroid polytope is bijective to its associated matroid and that every matroid polytope is a generalized permutohedra. The geometry of toric varieties are closely connected to matroids \cite{GelfandSerganova,GGMS}:
\begin{theorem}[{\cite[Lemma 1.4]{GGMS}}]
    The torus orbit of a point $p\in\Gr(k,n)$ is isomorphic to the toric variety defined by the matroid polytope of the representable matroid defined by the columns of the matrix of Steifel coordinates of $p$.
\end{theorem}

Another important property that is more on the algebraic side is that of \emph{normality}.
\begin{definition}
    The semigroup $\NN A$ is said to be \emph{normal} if $\NN A=\ZZ A\cap\RR_+ A$.
\end{definition}
Normality is satisfied by all closed torus orbits in a Grassmannian \cite{White1977}:
\begin{corollary}
    The closure of any torus orbit in a Grassmannian is projectively normal in its Pl\"ucker embedding.
\end{corollary}
The importance of normality in the study of Feynman integrals comes in that it guarantees that characteristic variety and dimension of the solution space of the generalized hypergeometric system $H_A(\beta)$ are independent of the parameters $\beta$ of the integral (that is, space-time dimension and propagator powers). Normality is in general stronger than this, what is actually of interest is that the toric ideal defined by $A$ should be \emph{Cohen-Macaulay} \cite{adolphson1994hypergeometric,Gelfand1993}, by a theorem of Hochster \cite{hochster1972rings}, normality implies the Cohen-Macaulay property.

Normality is connected to the GP property according to following theorem (cf. \cite[Fig. 5]{Dlapa:2023cvx}).
\begin{theorem}
    If $P$ is a generalized permutohedron and $A=\ZZ^n\cap P$, then $A$ is normal.
\end{theorem}

\begin{example}
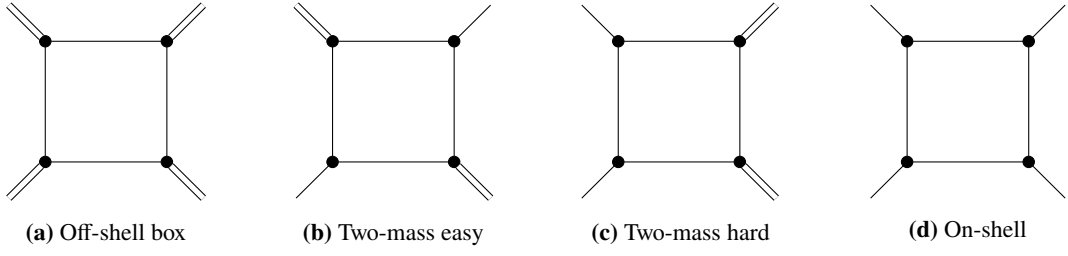
\begin{figure*}%
\captionsetup[subfigure]{justification=centering}
\newcommand{\xs}{.8}%
    \begin{subfigure}[t]{0.25\textwidth}
        \centering
         \begin{tikzpicture}[baseline=-\the\dimexpr\fontdimen22\textfont2\relax]
            \begin{feynman}
            \vertex [dot] (v1) at ( \xs, \xs) {};
            \vertex [dot] (v2) at (-\xs, \xs) {};
            \vertex [dot] (v3) at (-\xs,-\xs) {};
            \vertex [dot] (v4) at ( \xs,-\xs) {};

            \vertex (i1) at ( {1.6*\xs}, {1.6*\xs});
            \vertex (i2) at (-{1.6*\xs}, {1.6*\xs});
            \vertex (i3) at (-{1.6*\xs},-{1.6*\xs});
            \vertex (i4) at ( {1.6*\xs},-{1.6*\xs});
    \diagram*{
        (v1)[dot]--(v2)--(v3)--(v4)--(v1),
        (i1) -- [double,double distance=0.4ex](v1), (i2) -- [double,double distance=0.4ex](v2),(i3) -- [double,double distance=0.4ex](v3), (i4) -- [double,double distance=0.4ex](v4);
    };
    \end{feynman}
    \end{tikzpicture}
        \caption{Off-shell box}
        \label{fig:offshell}
    \end{subfigure}%
    \begin{subfigure}[t]{0.25\textwidth}
        \centering
         \begin{tikzpicture}[baseline=-\the\dimexpr\fontdimen22\textfont2\relax]
            \begin{feynman}
            \vertex [dot] (v1) at ( \xs, \xs) {};
            \vertex [dot] (v2) at (-\xs, \xs) {};
            \vertex [dot] (v3) at (-\xs,-\xs) {};
            \vertex [dot] (v4) at ( \xs,-\xs) {};

            \vertex (i1) at ( {1.6*\xs}, {1.6*\xs});
            \vertex (i2) at (-{1.6*\xs}, {1.6*\xs});
            \vertex (i3) at (-{1.6*\xs},-{1.6*\xs});
            \vertex (i4) at ( {1.6*\xs},-{1.6*\xs});
    \diagram*{
        (v1)[dot]--(v2)--(v3)--(v4)--(v1),
        (i1) -- (v1), (i2) -- [double,double distance=0.4ex](v2),(i3) -- (v3), (i4) -- [double,double distance=0.4ex](v4);
    };
    \end{feynman}
    \end{tikzpicture}
        \caption{Two-mass easy}
        \label{fig:twomasseasy}
    \end{subfigure}%
    \begin{subfigure}[t]{0.25\textwidth}
        \centering
        \begin{tikzpicture}[baseline=-\the\dimexpr\fontdimen22\textfont2\relax]
            \begin{feynman}
            \vertex [dot] (v1) at ( \xs, \xs) {};
            \vertex [dot] (v2) at (-\xs, \xs) {};
            \vertex [dot] (v3) at (-\xs,-\xs) {};
            \vertex [dot] (v4) at ( \xs,-\xs) {};

            \vertex (i1) at ( {1.6*\xs}, {1.6*\xs});
            \vertex (i2) at (-{1.6*\xs}, {1.6*\xs});
            \vertex (i3) at (-{1.6*\xs},-{1.6*\xs});
            \vertex (i4) at ( {1.6*\xs},-{1.6*\xs});
    \diagram*{
        (v1)[dot]--(v2)--(v3)--(v4)--(v1),
        (i1) -- [double,double distance=0.4ex](v1), (i2) -- (v2),(i3) -- (v3), (i4) -- [double,double distance=0.4ex]  (v4);
    };
    \end{feynman}
    \end{tikzpicture}
        \caption{Two-mass hard}
        \label{fig:twomasshard}
    \end{subfigure}%
    \begin{subfigure}[t]{0.25\textwidth}
        \centering
         \begin{tikzpicture}[baseline=-\the\dimexpr\fontdimen22\textfont2\relax]
            \begin{feynman}
            \vertex [dot] (v1) at ( \xs, \xs) {};
            \vertex [dot] (v2) at (-\xs, \xs) {};
            \vertex [dot] (v3) at (-\xs,-\xs) {};
            \vertex [dot] (v4) at ( \xs,-\xs) {};

            \vertex (i1) at ( {1.6*\xs}, {1.6*\xs});
            \vertex (i2) at (-{1.6*\xs}, {1.6*\xs});
            \vertex (i3) at (-{1.6*\xs},-{1.6*\xs});
            \vertex (i4) at ( {1.6*\xs},-{1.6*\xs});
    \diagram*{
        (v1)[dot]--(v2)--(v3)--(v4)--(v1),
        (i1) -- (v1), (i2) -- (v2),(i3) -- (v3), (i4) -- (v4);
    };
    \end{feynman}
    \end{tikzpicture}
        \caption{On-shell}
        \label{fig:onshell}
    \end{subfigure}
    \caption{Box diagrams with all internal masses equal to zero and some external legs being off-shell ($p^2\neq 0$, denoted by double lines) and some on-shell $(p^2=0)$.}
\label{fig:boxes tikz-feynman dashed}
\end{figure*}
We consider four different kinematic setups for the one-loop box integral with all internal masses equal to zero, see Fig. \ref{fig:boxes tikz-feynman dashed}.

The $A$-matrix for the off-shell box is
\begin{equation}
    A=\begin{pmatrix}
        1&1&1&1&0&0&0&0&0&0\\
        1&0&0&0&1&1&1&0&0&0\\
        0&1&0&0&1&0&0&1&1&0\\
        0&0&1&0&0&1&0&1&0&1\\
        0&0&0&1&0&0&1&0&1&1
    \end{pmatrix}
\end{equation}
which is precisely the vertices of the second hypersimplex $\Delta(2,5)$ and the toric variety $X_A=\mathbf{V}(I_A)$ is a Veronese-like embedding of $\PP^4$. This variety is isomorphic to the orbit closure of a generic point in the Grassmannian $\Gr(2,5)$ under the natural mapping of $(\CC^*)^5$. Also note that $\mathrm{conv}(A)$ is the basis polytope of the uniform matroid $U_{2,5}$. From this we know that $\NN A$ is normal (hence $I_A$ is Cohen-Macaulay) and $\conv(A)$ is a generalized permutohedron.

The three-mass box and two-mass easy, Fig. \ref{fig:twomasseasy}, correspond to sub matroid strata and thus normality and the GP property follows directly.

However, the two-mass hard (Fig. \ref{fig:twomasshard}), one-mass and on-shell box (Fig. \ref{fig:onshell}) do not correspond to any matroid strata. This means that $\conv(A)$ is not a matroid polytope so neither normality nor the GP property follows directly. From the results in \cite[Corollary 5.6]{Dlapa:2023cvx} it follows that $\NN A$ is normal, however, $\conv(A)$ is not a GP.  
\end{example}

Below we summarize some of the known results for GP and normality, we always use $A=\{1\}\times\mathrm{Supp}(\cG)$ with $\cG=\cU+\cF$ and $\newt[f]=\conv(\mathrm{Supp}(f))$ as the Newton polytope of $f$.
\begin{itemize}
    \item $\newt[\cU]$ is a matroid polytope, thus always a GP.
    \item For $m_e\neq0$ for all $e\in E$, then $\newt[\F]=\newt[\U]+\Delta(1,|E|)$ and thus always a GP. Moreover, if no cancellation between $\F_0$ and $\F_m$ occurs, then $A$ is normal \cite{TH22}.
    \item If no cancellation between $\cF_0$ and $\cF_m$ occurs, $p(V')^2\neq0$ for all $V'\subset V_{\mathrm{ext}}$ and every internal vertex is connected to an external vertex via a massive path, then $\newt[\cF]$ is a GP and $A$ is normal \cite{walther2022feynman} cf. \cite{Dlapa:2023cvx}.
    \item When $m_e=0$ for all $e\in E$ and $V=V_{\mathrm{ext}}$, then $\newt[\cF]$ is a matroid polytope and thus GP and $A$ is normal \cite{TH22}.
    \item When $m_e=0$ for all $e\in E$ and $p(V')^2\neq 0$ for all $V'\subset V_{\mathrm{ext}}$, then $\newt[\cF]$ is a matroid polytope and thus GP and $A$ is normal \cite{walther2022feynman} cf. \cite{Dlapa:2023cvx}.
\end{itemize}

\section{The program \ft}
The program \ft is available at\\

\noindent\fbox{\parbox{\textwidth}{
\href{https://github.com/michibo/feyntrop}{https://github.com/michibo/feyntrop}}}\\

It is a \texttt{C++} program with \texttt{Python} and \texttt{JSON} interface. It uses the contour deformation from section \ref{sec: contour deformation} and the sampling relies on the generalized permutohedron property, section \ref{sec: GP}.
\begin{example}[{\cite[Section 6.5]{Borinsky:2023jdv}}]
The following 5-point process with three massive external legs and a massive loop
    \begin{equation*}
        \begin{tikzpicture}[baseline=-\the\dimexpr\fontdimen22\textfont2\relax,scale=0.7]
        \begin{feynman}
            \vertex [dot,label={\(1\)}] (v1) at ( -1, 1) {};
            \vertex [dot,label={\(6\)}] (v2) at (1, 1) {};
            \vertex [dot,label=270:{\(5\)}] (v3) at (1,-1) {};
            \vertex [dot,label=270:{\(0\)}] (v4) at ( -1,-1) {};
            \vertex [dot,label={\(2\)}] (v5) at (3,1) {};
            \vertex [dot,label=270:{\(4\)}] (v6) at (3,-1) {};
            \vertex [dot,label=270:{\(3\)}] (v7) at (4,0) {};

            \vertex (i1) at ( -{1.8}, {1.8});
            \vertex (i2) at ({3.8}, {1.8});
            \vertex (i3) at ({3.8},-{1.8});
            \vertex (i4) at ( -{1.8},-{1.8});
            \vertex (i5) at (5.2,0);

            \vertex [circle] (o) at (0,0) {};
            
            \diagram*{
            (i1)--[scalar](v1)--(v2)--(v5)--(v7)--(v6)--(v3)--(v4)--(v1),
            (v2)--[scalar](v3),
            (i2)--[double,double distance=0.4ex](v5),
            (i5)--[double,double distance=0.4ex](v7),
            (i3)--[double,double distance=0.4ex](v6),
            (i4)--[scalar](v4),
            };
        \end{feynman}
    \end{tikzpicture}
    \end{equation*}
can be evaluated to percent precision to five orders in the dimensional regulator $\epsilon$ on a laptop in two seconds.
\end{example}

\bibliographystyle{JHEP}
\bibliography{Ref}

\providecommand{\href}[2]{#2}\begingroup\raggedright\begin{thebibliography}{10}

\bibitem{Heinrich:2020ybq}
G.~Heinrich, \emph{Collider physics at the precision frontier}, \href{https://doi.org/10.1016/j.physrep.2021.03.006}{\emph{Phys. Rept.} {\bfseries 922} (2021) 1--69}, [\href{https://arxiv.org/abs/2009.00516}{{\ttfamily 2009.00516}}].

\bibitem{Bern:2019nnu}
Z.~Bern, C.~Cheung, R.~Roiban, C.-H. Shen, M.~P. Solon and M.~Zeng, \emph{{Scattering Amplitudes and the Conservative Hamiltonian for Binary Systems at Third Post-Minkowskian Order}}, \href{https://doi.org/10.1103/PhysRevLett.122.201603}{\emph{Phys. Rev. Lett.} {\bfseries 122} (2019) 201603}, [\href{https://arxiv.org/abs/1901.04424}{{\ttfamily 1901.04424}}].

\bibitem{Dlapa:2020cwj}
C.~Dlapa, J.~Henn and K.~Yan, \emph{{Deriving canonical differential equations for Feynman integrals from a single uniform weight integral}}, \href{https://doi.org/10.1007/JHEP05(2020)025}{\emph{JHEP} {\bfseries 05} (2020) 025}, [\href{https://arxiv.org/abs/2002.02340}{{\ttfamily 2002.02340}}].

\bibitem{Karshenboim:2005iy}
S.~G. Karshenboim, \emph{{Precision physics of simple atoms: QED tests, nuclear structure and fundamental constants}}, \href{https://doi.org/10.1016/j.physrep.2005.08.008}{\emph{Phys. Rept.} {\bfseries 422} (2005) 1--63}, [\href{https://arxiv.org/abs/hep-ph/0509010}{{\ttfamily hep-ph/0509010}}].

\bibitem{zinn2021quantum}
J.~Zinn-Justin, \emph{Quantum field theory and critical phenomena}, vol.~171.
\newblock Oxford University Press, 2021.

\bibitem{Henn:2013pwa}
J.~M. Henn, \emph{{Multiloop integrals in dimensional regularization made simple}}, \href{https://doi.org/10.1103/PhysRevLett.110.251601}{\emph{Phys. Rev. Lett.} {\bfseries 110} (2013) 251601}, [\href{https://arxiv.org/abs/1304.1806}{{\ttfamily 1304.1806}}].

\bibitem{Liu:2022chg}
X.~Liu and Y.-Q. Ma, \emph{{AMFlow: A Mathematica package for Feynman integrals computation via auxiliary mass flow}}, \href{https://doi.org/10.1016/j.cpc.2022.108565}{\emph{Comput. Phys. Commun.} {\bfseries 283} (2023) 108565}, [\href{https://arxiv.org/abs/2201.11669}{{\ttfamily 2201.11669}}].

\bibitem{Hidding:2020ytt}
M.~Hidding, \emph{{DiffExp, a Mathematica package for computing Feynman integrals in terms of one-dimensional series expansions}}, \href{https://doi.org/10.1016/j.cpc.2021.108125}{\emph{Comput. Phys. Commun.} {\bfseries 269} (2021) 108125}, [\href{https://arxiv.org/abs/2006.05510}{{\ttfamily 2006.05510}}].

\bibitem{Armadillo:2022ugh}
T.~Armadillo, R.~Bonciani, S.~Devoto, N.~Rana and A.~Vicini, \emph{{Evaluation of Feynman integrals with arbitrary complex masses via series expansions}}, \href{https://doi.org/10.1016/j.cpc.2022.108545}{\emph{Comput. Phys. Commun.} {\bfseries 282} (2023) 108545}, [\href{https://arxiv.org/abs/2205.03345}{{\ttfamily 2205.03345}}].

\bibitem{Panzer:2014caa}
E.~Panzer, \emph{{Algorithms for the symbolic integration of hyperlogarithms with applications to Feynman integrals}}, \href{https://doi.org/10.1016/j.cpc.2014.10.019}{\emph{Comput. Phys. Commun.} {\bfseries 188} (2015) 148--166}, [\href{https://arxiv.org/abs/1403.3385}{{\ttfamily 1403.3385}}].

\bibitem{Borowka:2017idc}
S.~Borowka, G.~Heinrich, S.~Jahn, S.~P. Jones, M.~Kerner, J.~Schlenk et~al., \emph{{pySecDec: a toolbox for the numerical evaluation of multi-scale integrals}}, \href{https://doi.org/10.1016/j.cpc.2017.09.015}{\emph{Comput. Phys. Commun.} {\bfseries 222} (2018) 313--326}, [\href{https://arxiv.org/abs/1703.09692}{{\ttfamily 1703.09692}}].

\bibitem{Borinsky:2023jdv}
M.~Borinsky, H.~J. Munch and F.~Tellander, \emph{{Tropical Feynman integration in the Minkowski regime}}, \href{https://doi.org/10.1016/j.cpc.2023.108874}{\emph{Comput. Phys. Commun.} {\bfseries 292} (2023) 108874}, [\href{https://arxiv.org/abs/2302.08955}{{\ttfamily 2302.08955}}].

\bibitem{Panzer:2019yxl}
E.~Panzer, \emph{{Hepp's bound for Feynman graphs and matroids}}, \href{https://doi.org/10.4171/aihpd/126}{\emph{Ann. Inst. Henri Poincar\'{e} D} {\bfseries 10} (8, 2023) 31--119}, [\href{https://arxiv.org/abs/1908.09820}{{\ttfamily 1908.09820}}].

\bibitem{Borinsky:2020rqs}
M.~Borinsky, \emph{{Tropical Monte Carlo quadrature for Feynman integrals}}, \href{https://doi.org/10.4171/AIHPD/158}{\emph{Ann. Inst. Henri Poincar\'e Comb. Phys. Interact.} (8, 2023) }, [\href{https://arxiv.org/abs/2008.12310}{{\ttfamily 2008.12310}}].

\bibitem{Henn:2023vbd}
J.~M. Henn, J.~Lim and W.~J. Torres~Bobadilla, \emph{{First look at the evaluation of three-loop non-planar Feynman diagrams for Higgs plus jet production}}, \href{https://doi.org/10.1007/JHEP05(2023)026}{\emph{JHEP} {\bfseries 05} (2023) 026}, [\href{https://arxiv.org/abs/2302.12776}{{\ttfamily 2302.12776}}].

\bibitem{Balduf:2023ilc}
P.-H. Balduf, \emph{{Statistics of Feynman amplitudes in $\phi^4$-theory}},  \href{https://arxiv.org/abs/2305.13506}{{\ttfamily 2305.13506}}.

\bibitem{Arkani-Hamed:2022cqe}
N.~Arkani-Hamed, A.~Hillman and S.~Mizera, \emph{{Feynman polytopes and the tropical geometry of UV and IR divergences}}, \href{https://doi.org/10.1103/PhysRevD.105.125013}{\emph{Phys. Rev. D} {\bfseries 105} (2022) 125013}, [\href{https://arxiv.org/abs/2202.12296}{{\ttfamily 2202.12296}}].

\bibitem{Arkani-Hamed:2023lbd}
N.~Arkani-Hamed, H.~Frost, G.~Salvatori, P.-G. Plamondon and H.~Thomas, \emph{{All Loop Scattering as a Counting Problem}},  \href{https://arxiv.org/abs/2309.15913}{{\ttfamily 2309.15913}}.

\bibitem{Hannesdottir:2022bmo}
H.~S. Hannesdottir and S.~Mizera, \emph{{What is the i\ensuremath{\varepsilon} for the S-matrix?}}
\newblock SpringerBriefs in Physics. Springer, 1, 2023, \href{https://doi.org/10.1007/978-3-031-18258-7}{10.1007/978-3-031-18258-7}.

\bibitem{Lee2013}
R.~N. Lee and A.~A. Pomeransky, \emph{{Critical points and number of master integrals}}, \href{https://doi.org/10.1007/JHEP11(2013)165}{\emph{JHEP} {\bfseries 11} (2013) 165}, [\href{https://arxiv.org/abs/1308.6676}{{\ttfamily 1308.6676}}].

\bibitem{delaCruz:2019skx}
L.~de~la Cruz, \emph{{Feynman integrals as A-hypergeometric functions}}, \href{https://doi.org/10.1007/JHEP12(2019)123}{\emph{JHEP} {\bfseries 12} (2019) 123}, [\href{https://arxiv.org/abs/1907.00507}{{\ttfamily 1907.00507}}].

\bibitem{Klausen:2019hrg}
R.~P. Klausen, \emph{{Hypergeometric Series Representations of Feynman Integrals by GKZ Hypergeometric Systems}}, \href{https://doi.org/10.1007/JHEP04(2020)121}{\emph{JHEP} {\bfseries 04} (2020) 121}, [\href{https://arxiv.org/abs/1910.08651}{{\ttfamily 1910.08651}}].

\bibitem{Nilsson2013}
L.~Nilsson and M.~Passare, \emph{Mellin transforms of multivariate rational functions}, \href{https://doi.org/10.1007/s12220-011-9235-7}{\emph{J. Geom. Anal.} {\bfseries 23} (2013) 24--46}.

\bibitem{Berkesch2014}
C.~Berkesch, J.~Forsg{\aa}rd and M.~Passare, \emph{Euler-{M}ellin integrals and {$A$}-hypergeometric functions}, \href{https://doi.org/10.1307/mmj/1395234361}{\emph{Michigan Math. J.} {\bfseries 63} (2014) 101--123}.

\bibitem{Gelfand1986}
I.~M. Gel'fand, \emph{General theory of hypergeometric functions}, {\emph{Dokl. Akad. Nauk SSSR} {\bfseries 288} (1986) 14--18}.

\bibitem{Gelfand1987}
I.~M. Gel'fand, M.~I. Graev and A.~V. Zelevinski\u{\i}, \emph{Holonomic systems of equations and series of hypergeometric type}, {\emph{Dokl. Akad. Nauk SSSR} {\bfseries 295} (1987) 14--19}.

\bibitem{Gelfand1989}
I.~M. Gel'fand, A.~V. Zelevinski\u{\i} and M.~M. Kapranov, \emph{Hypergeometric functions and toric varieties}, \href{https://doi.org/10.1007/BF01078777}{\emph{Funktsional. Anal. i Prilozhen.} {\bfseries 23} (1989) 12--26}.

\bibitem{Gelfand1990}
I.~M. Gel'fand, M.~M. Kapranov and A.~V. Zelevinski\u{\i}, \emph{Generalized {E}uler integrals and {$A$}-hypergeometric functions}, \href{https://doi.org/10.1016/0001-8708(90)90048-R}{\emph{Adv. Math.} {\bfseries 84} (1990) 255--271}.

\bibitem{Gelfand1993}
I.~M. Gel'fand, M.~M. Kapranov and A.~V. Zelevinski\u{\i}, \emph{A correction to the paper ``hypergeometric functions and toric varieties''}, \href{https://doi.org/10.1007/BF01078854}{\emph{Funct. Anal. Appl.} {\bfseries 27} (1993) 295--295}.

\bibitem{eisenbud1996binomial}
D.~Eisenbud and B.~Sturmfels, \emph{Binomial ideals}, \href{https://doi.org/10.1215/S0012-7094-96-08401-X}{\emph{Duke Math. J.} {\bfseries 84} (1996) 1--45}.

\bibitem{gelfand2008discriminants}
I.~M. Gel'fand, M.~Kapranov and A.~Zelevinsk\u{\i}, \emph{Discriminants, resultants, and multidimensional determinants}.
\newblock Springer Science \& Business Media, 2008.

\bibitem{Postnikov2005}
A.~Postnikov, \emph{Permutohedra, associahedra, and beyond}, \href{https://doi.org/10.1093/imrn/rnn153}{\emph{Int. Math. Res. Not.} (2009) 1026--1106}.

\bibitem{GelfandSerganova}
I.~M. Gel'fand and V.~V. Serganova, \emph{Combinatorial geometries and the strata of a torus on homogeneous compact manifolds}, {\emph{Uspekhi Mat. Nauk} {\bfseries 42} (1987) 107--134, 287}.

\bibitem{GGMS}
I.~M. Gel'fand, R.~M. Goresky, R.~D. MacPherson and V.~V. Serganova, \emph{Combinatorial geometries, convex polyhedra, and {S}chubert cells}, \href{https://doi.org/10.1016/0001-8708(87)90059-4}{\emph{Adv. in Math.} {\bfseries 63} (1987) 301--316}.

\bibitem{White1977}
N.~L. White, \emph{The basis monomial ring of a matroid}, \href{https://doi.org/10.1016/0001-8708(77)90060-3}{\emph{Adv. in Math.} {\bfseries 24} (1977) 292--297}.

\bibitem{adolphson1994hypergeometric}
A.~Adolphson, \emph{Hypergeometric functions and rings generated by monomials}, \href{https://doi.org/10.1215/S0012-7094-94-07313-4}{\emph{Duke Math. J.} {\bfseries 73} (1994) 269--290}.

\bibitem{hochster1972rings}
M.~Hochster, \emph{Rings of invariants of tori, {C}ohen-{M}acaulay rings generated by monomials, and polytopes}, \href{https://doi.org/10.2307/1970791}{\emph{Ann. of Math. (2)} {\bfseries 96} (1972) 318--337}.

\bibitem{Dlapa:2023cvx}
C.~Dlapa, M.~Helmer, G.~Papathanasiou and F.~Tellander, \emph{{Symbol Alphabets from the Landau Singular Locus}}, \href{https://doi.org/10.1007/JHEP10(2023)161}{\emph{JHEP} {\bfseries 10} (2023) 161}, [\href{https://arxiv.org/abs/2304.02629}{{\ttfamily 2304.02629}}].

\bibitem{TH22}
F.~Tellander and M.~Helmer, \emph{{Cohen-Macaulay Property of Feynman Integrals}}, \href{https://doi.org/10.1007/s00220-022-04569-6}{\emph{{Commun. Math. Phys.}} {\bfseries 399} (2022) 1021--1037}, [\href{https://arxiv.org/abs/2108.01410}{{\ttfamily 2108.01410}}].

\bibitem{walther2022feynman}
U.~Walther, \emph{{On Feynman graphs, matroids, and GKZ-systems}}, \href{https://doi.org/10.1007/s11005-022-01614-2}{\emph{Lett. Math. Phys.} {\bfseries 112} (2022) 120}, [\href{https://arxiv.org/abs/2206.05378}{{\ttfamily 2206.05378}}].

\end{thebibliography}\endgroup

\end{document}